\documentclass[%
reprint,
superscriptaddress,
showpacs,preprintnumbers,
 amsmath,amssymb,
aps,
prb,
floatfix,
]{revtex4-1}

\usepackage{epsfig}
\usepackage[thinspace]{SIunits}
\usepackage{xspace}
\usepackage{graphicx}
\usepackage{dcolumn}
\usepackage{bm}
\usepackage{color}
\usepackage{hyperref}
\hypersetup{
    unicode=false,          
    pdftoolbar=true,        
    pdfmenubar=true,        
    pdffitwindow=false,     
    pdfstartview={FitH},    
    pdftitle={My title},    
    pdfauthor={Author},     
    pdfsubject={Subject},   
    pdfcreator={Creator},   
    pdfproducer={Producer}, 
    pdfkeywords={keyword1} {key2} {key3}, 
    pdfnewwindow=true,      
    colorlinks=true,       
    linkcolor=blue,          
    citecolor=blue,        
    filecolor=blue,      
    urlcolor=blue,       
    plainpages=false        
}

\begin{document}

\title{Charge density wave modulation and gap measurements in CeTe$_3$}

\author{U. Ralevi\'c}
\affiliation{Center for Solid State Physics and New Materials, Institute of Physics, University of Belgrade, Pregrevica 118, 11080 Belgrade, Serbia}

\author{N. Lazarevi\'c}
\affiliation{Center for Solid State Physics and New Materials, Institute of Physics, University of Belgrade, Pregrevica 118, 11080 Belgrade, Serbia}

\author{A. Baum}
\affiliation{Walther Meissner Institut, Bayerische Akademie der Wissenschaften, 85748 Garching, Germany}

\author{H.-M. Eiter}
\affiliation{Walther Meissner Institut, Bayerische Akademie der Wissenschaften, 85748 Garching, Germany}

\author{R.~Hackl}
\affiliation{Walther Meissner Institut, Bayerische Akademie der Wissenschaften, 85748 Garching, Germany}

\author{P. Giraldo-Gallo}
\affiliation{Physics and Applied Physics, Stanford University, Stanford, California 94305, USA}

\author{I. R. Fisher}
\affiliation{Physics and Applied Physics, Stanford University, Stanford, California 94305, USA}

\author{C. Petrovic}
\affiliation{Condensed Matter Physics and Materials Science Department, Brookhaven National Laboratory, Upton, New York 11973, USA}

\author{R. Gaji\'c}
\affiliation{Center for Solid State Physics and New Materials, Institute of Physics, University of Belgrade, Pregrevica 118, 11080 Belgrade, Serbia}

\author{Z. V. Popovi\'c}
\affiliation{Center for Solid State Physics and New Materials, Institute of Physics, University of Belgrade, Pregrevica 118, 11080 Belgrade, Serbia}

\date{\today}

\begin{abstract}
  We present a study of charge density wave (CDW) ordering in CeTe$_3$ at room temperature using a scanning tunneling microscope and Raman spectroscopy. Two characteristic CDW ordering wavevectors obtained from the Fourier analysis are assessed to be $|{\bf \textbf{c}}^\ast-{\bf q}|=4.19\,{\rm nm}^{-1}$ and $|{\bf q}|=10.26\,{\rm nm}^{-1}$ where $|{\bf c}^\ast| = 2\pi/c$ is the reciprocal lattice vector. The scanning tunneling spectroscopy measurements, along with inelastic light (Raman) scattering measurements, show a CDW gap $\Delta_{\rm max}$ of approximately 0.37\,eV. In addition to the CDW modulation, we observe an organization of the Te sheet atoms in an array of alternating V- and N- groups along the CDW modulation, as predicted in the literature.
\end{abstract}


\maketitle

\section{Introduction}

Charge density waves (CDW) have been a subject of considerable interest in condensed matter physics for many decades.\cite{Gruner:1994} Originally expected and observed in one-dimensional (1D) systems, CDW formation was found also in many 2D materials where the ordered state remains metallic in most of the cases. Occasionally, the CDW phase is in close proximity to other phases such as superconductivity.\cite{Ghiringhelli:2012,Maple:2012}

The rare-earth tritellurides\cite{DiMasi:1995} $R$Te$_3$ are excellent model systems for systematic studies of the underlying physics for the wide range of tunable parameters such as transition temperature $T_{\rm CDW}$, Fermi surface shape, $c$-axis coupling, anisotropic or weak versus strong electron-phonon coupling.\cite{Yao:2006,Ru:2006,Sacchetti:2009,Sacchetti:2007,Malliakas:2005,Malliakas:2006,Ru:2008a,Eiter:2013} These compounds have an orthorhombic crystal structure ($Cmcm$ space group) commonly thought of as being 'weakly' orthorhombic, i.e. tetragonal with a 2-fold reduced symmetry due the very subtle difference of the in-plane crystal axes.\cite{Yao:2006} For the layered quasi two-dimensional (2D) material with the reduced tetragonal symmetry, the CDW ground state can either be bidirectional (checkerboard) or unidirectional (stripes).\cite{Yao:2006} High resolution x-ray diffraction,\cite{Malliakas:2005, Malliakas:2006,Ru:2008a} angle resolved photoemission spectroscopy (ARPES)\cite{Brouet:2004,Schmitt:2008,Moore:2010} and femtosecond pump-probe spectroscopy\cite{Yusupov:2008} showed that lighter rare-earth tritellurides (i.e. $R=\rm{La\dots Tb}$) host an unidirectional incommensurate CDW well above room temperature. For the heavy rare-earth tritellurides (i.e. $R=\rm{Dy\dots Tm}$) the upper transition temperature $T_{\rm CDW1}$ is below room temperature, and another transition having an orthogonal ordering vector occurs at $T_{\rm CDW2}<T_{\rm CDW1}$.

Scanning tunneling microscopy (STM) studies of $R$Te$_3$, performed at low temperatures,\cite{Fang:2007,Tomic:2009} show that the CDW modulations within the Te square lattice are incommensurate for the lighter $R$Te$_3$ compounds. In addition, using STM on TbTe$_3$, Fang \textit{et al.}\cite{Fang:2007} demonstrated a possible dimerization of the atoms in the Te plane. Slight modification of the atomic arrangement is not unexpected for systems exhibiting charge density waves. In fact, it was already theoretically predicted that the atoms in a square lattice have tendency toward different distortion phases.\cite{Hoffmann:1987} For CeTe$_3$ electron diffraction experiments suggest structural distortions in the Te sheets, recognised as N and V shaped groups of Te atoms, representing another broken symmetry of the crystal lattice.\cite{Patschke:2002,Malliakas:2005} This phenomenon was further studied by Kim \textit{et al.}\cite{Kim:2006} applying an atomic pair distribution function (PDF) analysis of the x-ray data. The distance between nearest neighbour Te-Te atoms therein was found to have a bimodal distribution indicating the presence of the aforementioned structural distortions. The occurrence of the N and V patterns in the Te sheets is, however, inconsistent with the present STM data, and it is, therefore, an open issue. Moreover, there is a discrepancy between the gap values derived from ARPES\cite{Brouet:2008} and a recent tunnelling study.\cite{Tomic:2009}

In this article, we reveal the CDW modulation in CeTe$_3$ at room temperature and demonstrate the periodic pattern of Te sheet atoms grouped in N and V shapes. We assess the maximal CDW gap $\Delta_{\rm max}$  by both scanning tunneling spectroscopy (STS) and Raman scattering measurements and find it to be close 0.37\,eV.

\section{Experimental}

Single crystals of CeTe$_3$ were grown by slow cooling of a binary melt. The details are published elsewhere.\cite{Ru:2006} The samples for STM measurements were prepared at Brookhaven National Laboratory, whereas those for Raman scattering at Stanford University. Because the compound oxidizes in ambient air, the crystals were cleaved with adhesive tape in order to remove the oxidized layer.

The STM measurements were done in ultra-high vacuum at room temperature, using a commercial OMICRON UHV STM system. Topography imaging was done in constant current mode with the bias voltages in range of 0.1-0.9\,V applied to the tip. Tilting error compensation, Fourier noise filtering, and scanner distortions removal were done using Gwyydion image processing program.\cite{Necas:2012} Every STM topograph and its corresponding current image are displayed with their raw Fourier transform, used for the filtering procedure. STS measurements were done in point probe mode at different locations on the sample surface, also at room temperature. The I-V curves were made for bias voltages ranging from -0.5\,V to 0.5\,V, with a resolution of 0.02\,V, which is comparable to the thermal noise of about 0.026\,eV.

For the Raman scattering experiments we have used the line at 458\,nm of an Ar ion laser as an excitation source. The absorbed laser power ranged from 1 to 2\,mW to keep the local heating below 5\,K in the $50\times 100\,\mu{\rm m}^2$-sized focus. The spectra were measured with a resolution of 2.5\,cm$^{-1}$ at low energy and 20\,cm$^{-1}$ at high energy. The Raman response $R\chi^{\prime\prime}(\Omega)$ is then obtained by dividing the measured spectra by the thermal Bose factor. We present symmetry-resolved spectra, obtained from linear combinations of spectra measured at the main polarization configurations,\cite{Muschler:2010a} in order to separate out the $A_{2g}$ contribution which, in the non-resonant case, is insensitive to the carrier (particle-hole) excitations we are interested in.

\section{Results and discussions}

\subsection{STM topography measurements}\label{S3_2}

\begin{figure}[htb!]
  \centering
  \includegraphics{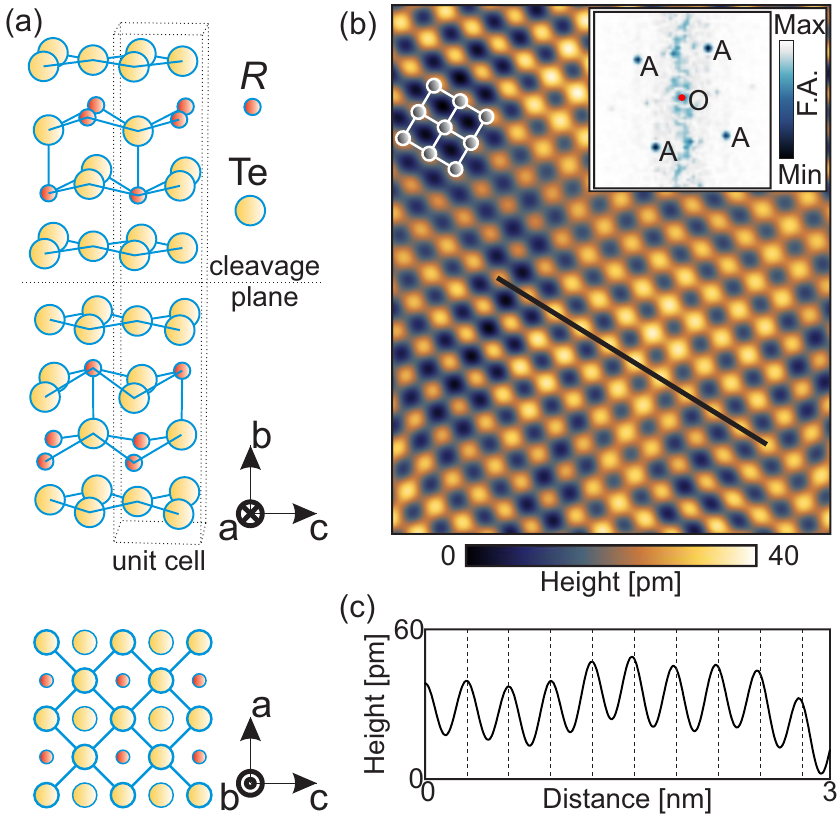}
  \caption{(Color online) (a) Crystal structure of CeTe$_3$. The lower part shows a view on the top Te layer connected by blue lines oriented at 45$^\mathrm{o}$ with respect to the $a$ and $c$ axes. The atoms of the $R$-Te layer are not connected. (b) Filtered STM topograph obtained for $V_b$ =0.8\,V and $I_s$ =0.6\,nA. The inset shows corresponding Fourier transform (F.A. stands for Fourier Amplitude); (c) Profile along the black line in panel (b).
  \label{fig_1}
  }
\end{figure}

CeTe$_3$ crystals are layered quasi-2D materials with orthorhombic crystal structure that adopts the $Cmcm$ space group symmetry.\cite{Norling:1966} The layers are composed of a corrugated CeTe slab sandwiched between two planar Te sheets and connected together via weak van der Waals forces. Figure \ref{fig_1} (a) shows the crystal structure of CeTe$_3$ with the unit cell indicated by dashed lines. For the $Cmcm$ space group the $a$ and $c$ are the in-plane crystal axes and $b$ is the long crystal axis.

The planar Te sheets have a nearly perfect square lattice with one Te atom in the unit cell, as opposed to full crystal having two Te atoms per unit cell. As an example, in Fig. \ref{fig_1} (b) the Te square lattice of CeTe$_3$ obtained by setting the bias voltage, $V_b$, to 0.8\,V and the set-point current, $I_s$, to 0.6\,nA, is shown. The corresponding raw Fourier transform is given in the inset. Four distinct peaks, labeled as A, form the reciprocal lattice of the atoms in the Te sheets, and are approximately $2\pi/a_0$=20.43\,nm$^{-1}$ from the origin, denoted as O. The average distance between the neighbouring Te atoms can be straightforwardly assessed from the position of the A peaks, i.e. $a_0=2\pi$/20.43\,nm$^{-1}$=0.3075\,nm. However, profile taken along the solid (black) line in Fig. \ref{fig_1} (b) reveals that the distances between the maxima, representing the Te atoms, do not appear at a constant separation of $2\pi$/20.43\,nm$^{-1}$=0.3075\,nm indicated by dashed lines. The variation of the Te-Te distances is very subtle and can not be assessed from the full width at half maximum of the A peaks (see the inset in Fig. \ref{fig_1} (b)). Therefore, the Fourier transform provides the average interatomic distances indicating an ideal Te square lattice, whereas the analysis of the real space images yields irregular Te-Te distances.

\begin{figure}[htb!]
  \centering
  \includegraphics[width=1.0\columnwidth]{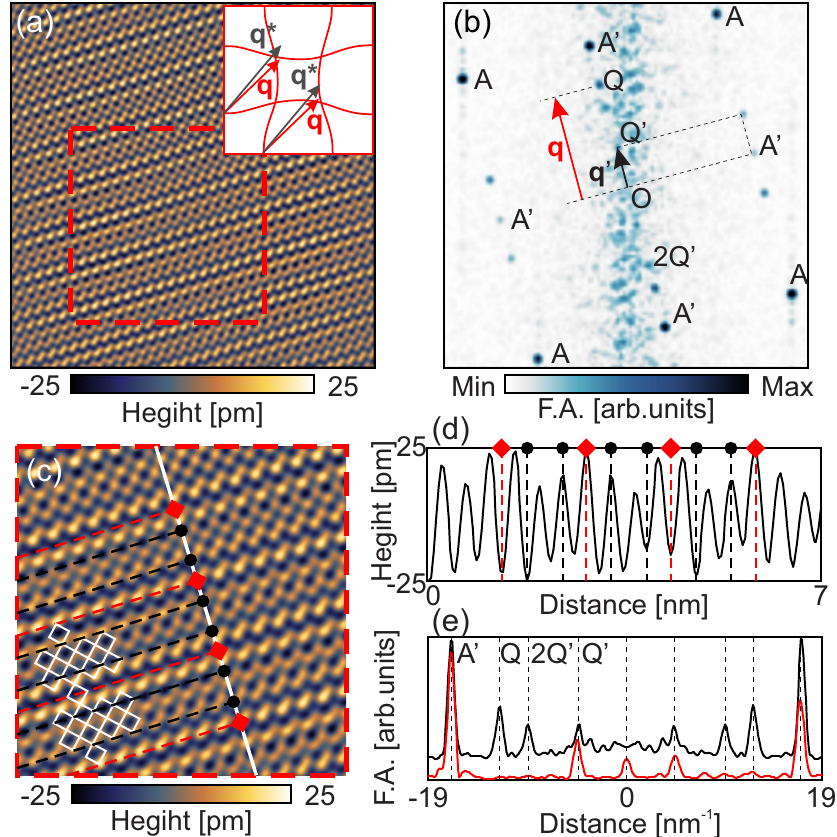}
  \caption{(Color online) (a) Flittered STM topography obtained for $V_b=0.2$\,V and $I_s=0.6$\,nA. The inset shows a sketch of Fermi surface and the possible nesting vectors; (b) The corresponding Fourier transform; (c) Magnified part of image in (a) showing the CDW modulation; (d) Profile along the white line in panel (c); (e) Profile along the main and side diagonals of the Fourier image in panel (b). F.A. stands for Fourier Amplitude.
  }
  \label{fig_2}
\end{figure}

Given the weak hybridization between the CeTe slab and the Te sheets,\cite{Yao:2006,Kikuchi:1998} the electronic properties of $R$Te$_3$ are predominantly determined by the planar Te sheets. Having a weak interaction along the $b$ axis, these compounds are often considered quasi two-dimensional (2D) systems. The main contribution to the CDW comes from the states close to the Fermi level. These states mainly derive from in-plane $p_x$ and $p_z$ orbitals with 5/8 filling due to the underlying CeTe layer, which donates an electron shared between two atoms in the upper and lower Te sheet. The out-of-plane $p_y$ orbital is lower in energy than the in-plane $p_x$ and $p_z$ orbitals and is pushed below the Fermi level. Now, assuming an isolated Te sheet having an ideal square lattice, whose $p$ orbitals are filled as explained, one could calculate the Fermi surface\cite{Yao:2006,Eiter:2013,Eiter:2014b} within the corresponding Brillouin zone similar to the one sketched in the inset of Figure \ref{fig_2} (a) with solid (red) lines. In the scope of this approximation, there are two possibilities for the nesting vectors, ${\bf q}$ and ${\bf q}^\ast$\cite{Johannes:2008} which are shown in Fig. \ref{fig_2} (a). Assuming a nesting-driven instability, while neglecting the hopping between $p_x$ and $p_z$ orbitals, allows the system to choose between the two nesting vectors depending on the strength of the electron-phonon coupling: ${\bf q}$ is favoured when the coupling is strong.\cite{Yao:2006} It was recently shown,\cite{Eiter:2013} however, that the inclusion of the interaction between $p_x$ and $p_z$ modifies the Fermi surface, thus, enabling large electron-phonon coupling near the band degeneracy points and renormalization of the phonon frequency only at the ${\bf q}$, rather than at ${\bf q}^\ast$.

Fig. \ref{fig_2} (a) shows the topography of CeTe$_3$ with clearly visible CDW modulation. This image is obtained for $V_b$=0.2\,V and $I_s$=0.6\,nA. The corresponding Fourier image is given in Fig. \ref{fig_2} (b). As before, there are four peaks representing the Te sheet atoms. They are labeled by A. Peaks labeled by A$^\prime$ correspond to the additional periodicity arising from the inequality of ${\bf a}$ and ${\bf c}$ lattice vectors which is a consequence of the orthorhombic crystal structure. In other words, these peaks should exist for any set of tunneling parameters for which the tunneling between the tip and the Te planes is possible. In fact, the A$^\prime$ peaks exist in Fig. \ref{fig_1} (b) as well, but are barely visible since the inequality between ${\bf a}$ and ${\bf c}$ is very subtle. The intensity of the A$^\prime$ peaks in Fig. \ref{fig_2} (b) is ''enhanced'' by the additional periodicity of the underlaying Ce atoms, whose $f$ states have significant contribution to the tunneling current\cite{Tomic:2009} when the tip is close to the surface, i.e. for lower bias voltages. Accordingly, we assign ${\bf c}^\ast$ and ${\bf a}^\ast$ wavevectors to the A$^\prime$ peaks, where $|{\bf c}^\ast|=2\pi/|{\bf c}|$ and $|{\bf a}^\ast|=2\pi/|{\bf a}|$. The ${\bf c}^\ast$ connects the origin O and the A$^\prime$ peak on the main diagonal, and the ${\bf a}^\ast$ is perpendicular to ${\bf c}^\ast$. We assess the magnitude of the two wavevectors to be around 14.4\,nm$^{-1}$.

Apart from A and A$^\prime$ peaks, the Fourier amplitude image in Fig. \ref{fig_2} (b) exhibits new features in comparison to the one shown in Fig. \ref{fig_1} (b). The peak located below the A$^\prime$ peak on the main diagonal, Q, corresponds to the ${\bf q}$ nesting vector, while the Q$^\prime$ peak corresponds to the ${\bf q}^\prime$ vector. Following the work in Ref. \onlinecite{Eiter:2013} we set the ${\bf q}$ as the dominant CDW ordering vector, which further gives ${\bf q}^\prime={\bf c}^\ast-{\bf q}$. The magnitudes of ${\bf q}$ and ${\bf c}^\ast-{\bf q}$ vectors are 10.26\,nm$^{-1}$ and 4.19\,nm$^{-1}$, respectively. The related wavelengths are $\sim$0.61\,nm (distance between (black) circles in Figs. \ref{fig_2} (c) and (d)) and $\sim$1.5\,nm (distance between (red) squares in Figs. \ref{fig_2}(c) and (d)), respectively. Displayed in Fig. \ref{fig_2} (e) are the profiles along main and side diagonal of the Fourier image, showing the exact position of the mentioned peaks. The Q peak is more intensive than the Q$^\prime$ peak which is barely visible on the main diagonal, but clearly visible on the side diagonal. The absence of the Q or Q$^\prime$ peaks on the main diagonal in Fig. \ref{fig_1} (b), is due to the small, practically negligible, overlapping of the CDW wavefunction, which decays exponentially out of Te sheet, and the tip wavefunction. In fact, significantly reduced overlapping between the two wavefunctions for $V_b= 0.8$\,V results in a reduced CDW current signal comparable to the noise level.

\begin{figure}[htb!]
  \centering
  \includegraphics{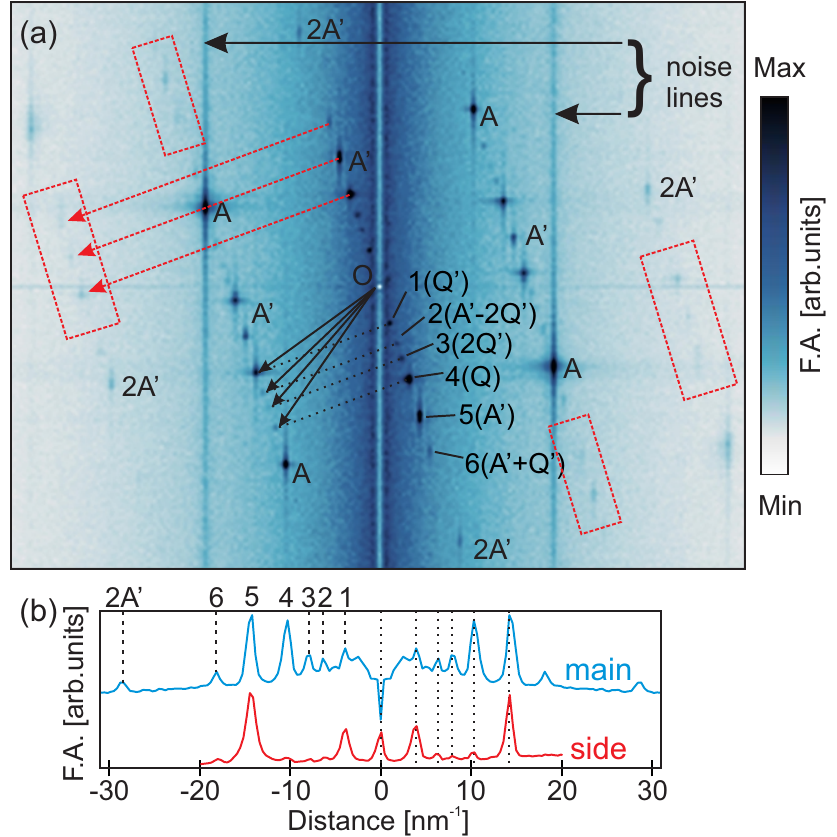}
  \caption{(Color online) (a) Averaged Fourier transform obtained by summing 34 Fourier transforms which are similar to the one shown in Fig. \ref{fig_1} (b); (b) Profile along the main and side diagonals. Red rectangles mark the areas within which are the Fourier peaks representing higher order harmonics. F.A. stands for Fourier Amplitude.
  }
  \label{fig_3}
\end{figure}

In addition to Q and Q$^\prime$, another peak, which we label as 2Q$^\prime$, exists on the main diagonal of the Fourier image in Fig. \ref{fig_2} (b). It is located at 8.38\,nm$^{-1}$ from the origin, which is exactly twice the distance between the origin and the Q$^\prime$ peak. The wave vector connected to this peak is, thus, exactly two times larger than the ${\bf c}^\ast-{\bf q}$. Both Q$^\prime$ and 2Q$^\prime$ are, in fact, due to the wavevector mixing effect which occurs between the ${\bf q}$ and ${\bf c}^\ast$.\cite{Fang:2007} However, the full mixing effect is not entirely visible in Fig. \ref{fig_2} (b) because the exerted noise masks the rest of the prominent mixing peaks. In Fig. \ref{fig_3} (a), which has been obtained by averaging 34 consecutive Fourier images, this effect becomes more clear. As it can be seen, additional peaks arise by improving the signal-to-noise ratio, both on the main and the side diagonal. Numbers 2 and 6 in Fig. \ref{fig_3} (a) denote two new peaks on the main diagonal specified by $2{\bf q}-{\bf c}^\ast$ and $2{\bf c}^\ast-{\bf q}$ vectors, respectively. The side diagonal has four peaks corroborating the afore mentioned effect. Their wavevectors, ${\bf a}^\ast+{\bf c}^\ast-{\bf q}$, ${\bf a}^\ast+2{\bf q}-{\bf c}^\ast$, ${\bf a}^\ast+2{\bf c}^\ast-2{\bf q}$ and ${\bf a}^\ast+{\bf q}$ are shown by arrows in Fig. \ref{fig_3} (a). Finally, Fig. \ref{fig_3} (b) displays line cuts along the main and side diagonals in order to illustrate that every peak on the main diagonal, except A$^\prime$+Q$^\prime$ and 2A$^\prime$, has its ''copy'' on the side diagonal.

\begin{figure}[htb!]
 \centering
 \includegraphics{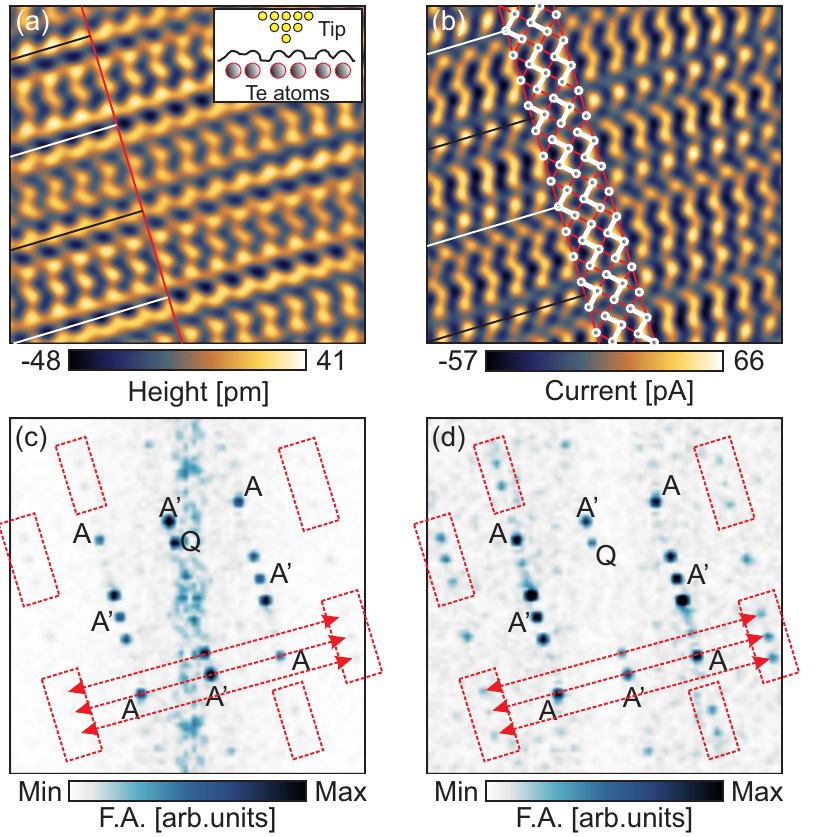}
 \caption{\label{fig_4} (Color online) (a) Topography and (b) current images obtained for $V_b$=0.1\,V and $I_s$=0.7\,nA and (c), (d)their corresponding Fourier images. Red rectangles mark the areas within which are the Fourier peaks representing higher order harmonics. F.A. stands for Fourier Amplitude.
 }
\end{figure}

As noted before, Te sheets are unstable and prone to the CDW. The superspace crystallographic analysis done in Ref. \onlinecite{Malliakas:2005} indicated that the distances between neighboring Te atoms in the planar sheets vary around the average of 0.3035\,nm in a systematic manner. The total ordering can be viewed as alternate sequences of V and N groups of Te sheet atoms in the direction of the crystallographic $c$ axis (along the CDW modulation). Analyzing Figs. \ref{fig_1} (b) and \ref{fig_1} (c) we have already noted that the interatomic distances between neighboring Te atoms vary around the average of 0.3075\,nm, which is slightly higher that the one given in Ref. \onlinecite{Malliakas:2005}. However, since the distance variations are rather subtle we were not able to confirm the proposed ordering by examining images similar to those in Fig. \ref{fig_1} (b). Therefore, we have chosen the tunneling parameters, $V_b$ = 0.1\,V and $I_s$ = 0.7\,nA, for which the tip is even closer to the surface than in the case shown in Fig. \ref{fig_2} (a). In this way, the close proximity of a sharp tip to the surface atoms would result in a convolution image such that the atoms being further away from each other would be resolved better than the atoms which are closer to each other (see the sketch in the inset of Figure \ref{fig_4} (a)).

The corresponding topography image is shown in Fig. \ref{fig_4} (a). Next to it (Fig. \ref{fig_4} (b)) is an image formed by measuring the changes of the tunneling current (with respect to $I_s$) before the feedback moves the piezo scanner in order to maintain the predefined current setpoint. Below them, in Figs. \ref{fig_4} (c) and \ref{fig_4} (d), shown are the corresponding Fourier images.

The topography image in Fig. \ref{fig_4} (a) clearly displays the CDW modulation stripes, as expected, with strange looking shapes in between the stripes. On the other hand, the current image in Fig. \ref{fig_4} (b) displays N and V groups of Te sheet atoms, indicated by higher current values, which have a repeating pattern matching the CDW modulation stripes in Fig. \ref{fig_4} (a). The observed patterns in both topography and current images may originate from the convolution of a damaged (not very sharp) tip and the surface or they may represent a local modification of the surface. That being the case, one would expect to observe new peaks in the Fourier images at the positions defined by wavevectors which are not represented as linear combinations of the lattice and the modulation wavevectors. However, the corresponding Fourier images in Figs. \ref{fig_4} (c) and \ref{fig_4} (d) have almost all of the characteristic peaks seen in Fig. \ref{fig_2} (b), i.e. A, A$^\prime$, Q and Q$^\prime$, without any new features. The absence of the 2Q$^\prime$ peak off the main diagonals in Figs. \ref{fig_4} (c) and \ref{fig_4} (d) is a consequence of applying Fourier transform to the images scanned over a small area and, off course, a consequence of low signal-to-noise ratio, rather than a consequence of the modification of the carrier density modulation. It is, therefore, natural to assume that only carrier density modulation, lattice features and, as explained, their interplay have an impact on the formation of the image in Figs. \ref{fig_4} (a) and \ref{fig_4} (b), and that the effect of the convolution of the surface with a ''bad'' tip can be ruled out.

Having in mind that the Fourier images in Figs. \ref{fig_4} (c) and \ref{fig_4} (d) have the same periodic components, one can conjecture that different motifs observed in the topography and current images originate from the different intensities of the Fourier peaks. In fact, the strong CDW stripe motif in the topography image is due to the fact that the Q peaks in Fig. \ref{fig_4} (c) have higher intensity than the A peaks. Oppositely, the current image does not have strong CDW stripe motif because the Q peaks have significantly lower intensity than the rest of the peaks in Fig. \ref{fig_4} (d). On the other hand, the shapes in Fig. \ref{fig_4} (a) and the N and V shapes in Fig. \ref{fig_4} (b) are mainly determined by the intensity of the Q$^\prime$, A and A$^\prime$ peaks.

The ''modulation'' features on the side diagonals of the Fourier images may exist due to both carrier density modulation and structural modulation (rearrangement) of the Te sheet atoms. If we were to assume that there is no structural modulation, than the intensity of the
aforementioned peaks would be determined by the strength of the tunneling signal from the carrier density modulation components with the corresponding periodicity. If the structural modulation is considered without the carrier density modulation, than the intensity of the related peaks would be determined by the magnitude of the atom displacements. Larger displacements would yield Fourier peaks with high intensity, but would also introduce asymmetry in the intensity of the ''modulation'' peaks across the side diagonal. Such asymmetry is seen as unequal intensity of the Q$^\prime$ peaks on the side diagonals in Figs. \ref{fig_4} (c) and \ref{fig_4} (d). Similar asymmetry of the Q$^\prime$ peak intensities can be observed in Fig. \ref{fig_2} (b). The Q$^\prime$ peaks on side diagonals in Fig. \ref{fig_2} (b) are, however, weaker in comparison to A and Q peaks, which is why the corresponding topography looks like a square Te lattice to which the carrier density modulation and its components are added. Therefore, the structural modulation seen in Fig. \ref{fig_4} (b) most probably exists also in Figs. \ref{fig_2} (b) and \ref{fig_4} (a) but it is masked by contribution of the other periodic components included in the formation on these images.

Finally, further examination of Fourier images in Figs. \ref{fig_3} (a), \ref{fig_4} (c) and \ref{fig_4} (d) indicates the existence of higher order harmonics (marked by red rectangles in the mentioned figures) which appear to be located at the positions corresponding to the linear combinations of the lattice and the modulation wavevectors. Hence, one could speculate that these higher order harmonics might be an indication of a structural modulation occurring in the Te sheets. However, the intensity of these peaks is very low, especially in Figs. \ref{fig_3} (a) and \ref{fig_4} (c), and comparable to the noise level. Unfortunately, we find that takeing larger scans than those shown in Figs. \ref{fig_4} (a) and \ref{fig_4} (b) does not improve signal-to-noise ratio for the higher harmonics at room temperature, but rather leads to an unstable operation in which the topography and the current images change their appearance after a certain time of scanning. This leaves the existence of higher harmonics as an open question.

\subsection{CDW gap measurement}
\subsubsection{STS measurements}
The formation of energy gaps is a general characteristic of the CDW systems. In these systems a broken symmetry ground state (CDW state) is stabilized via electron-electron or electron-phonon interactions or both, by creating electron-hole pairs with the ordering vector selected as explained in Section \ref{S3_2}. This, in turn, leads to the energy gap formation in the regions of the Fermi surface connected by the preselected ordering vector, thus, lowering the energy of the system. For $R$Te$_3$ family the Fermi surface is only partially gapped. Here we use STS in order to assess the CDW gap of CeTe$_3$ at room temperature.

\begin{figure}[htb!]
  \centering
  \includegraphics{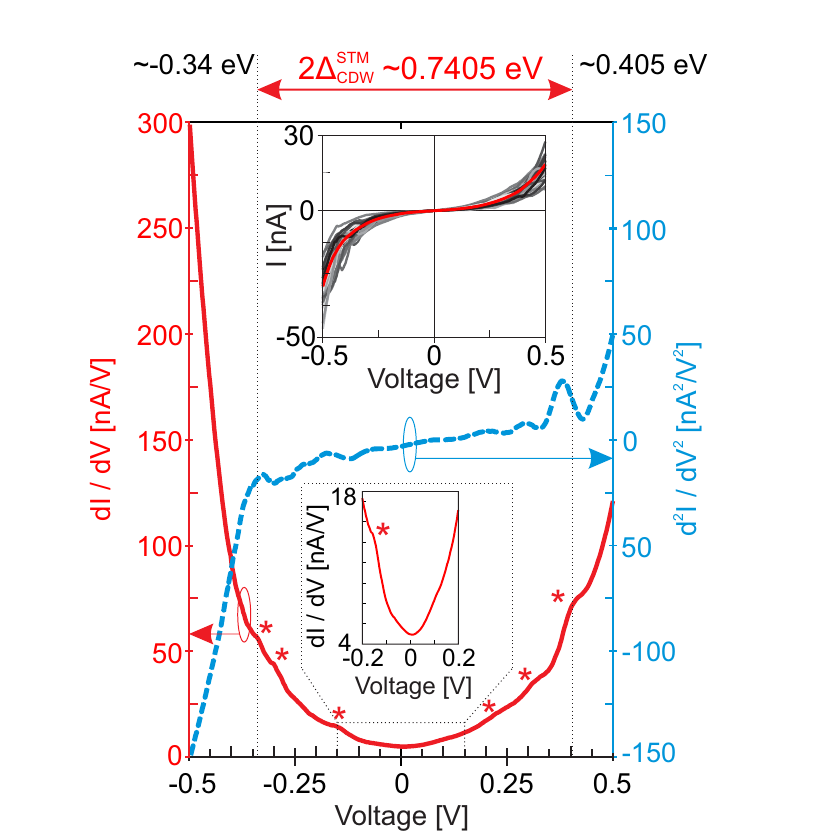}
  \caption{(Color online) Averaged $dI/dV$ curve obtained by repeated measurements at different positions on the sample and its first derivative. The upper inset shows all $I-V$ curves taken into account. The lower inset shows magnification of the $dI/dV$ curve marked by square with dashed borders. The red asterisks indicate the kinks.
  }
  \label{fig_5}
\end{figure}

Figure \ref{fig_5} (a) shows an averaged $dI/dV$ versus $V$ (red curve) and its first derivative, $d^2I/dV^2$ (blue curve). In the upper inset shown are the $I-V$ curves collected at different locations on the sample by performing ten consecutive measurements at each location. The averaging procedure was done due to the spatial variation on $I-V$ curves at random points, including ones above and in between the Te sheet atoms. The asymmetry of the $dI/dV$ curve originates from the fact that, apart from the in-plane $p_x$ and $p_z$ orbitals, the $p_y$ orbitals are probed as well.\cite{Fang:2007} Removal of electrons from $p_y$, which occurs at negative voltages, is more favourable than the adding of electrons at positive voltages since the $p_y$ orbitals are below the Fermi level. Therefore, the conductance, i.e. $dI/dV$, will be higher for negative than for positive voltages and, consequently, the $dI/dV$ curve would be asymmetric. This type of $dI/dV$ asymmetry is common among systems exhibiting CDW.\cite{Fang:2007, Tomic:2009, Machida:2013, Soumyanarayanan:2013, Flicker:2015} The V-shaped $dI/dV$ curve with a depressed DOS inside the gap and a finite conductance at zero bias suggest a partially gapped Fermi surface, as it was noted before. The V-shape becomes apparent after magnification of the $dI/dV$ inside the gap (see the lower inset in Fig. \ref{fig_5}).

The spatial variation of $I-V$ curves, in the upper inset, is probably the reason why the average $dI/dV$ exhibits subtle kinks marked by (red) stars in Fig. \ref{fig_5}. For each kink in the $dI/dV$ curve one can observe a slope in the corresponding $d^2I/dV^2$ curve which, indeed, corroborates that the kinks exist. These variations are likely due to involvement of the tip states, due some other effect such as zero bias anomaly, \cite{WangC:1990} or even maybe due to the existence of the shadow gaps in CeTe$_3$.\cite{Brouet:2004} Similar structures can be observed in Ref. \onlinecite{Tomic:2009}. Unfortunately, we can not claim with certainty what they represent nor determine their origin. Exceptions are the two features at $\sim-0.34$\,eV and $\sim0.405$\,eV, additionally marked by dashed lines, which correspond to the most prominent slopes of $d^2I/dV^2$ and which seem to be present in the most of the measurements. In fact, they represent the bands edges rising above and below the Fermi level, and, therefore, correspond to the CDW gap which originates from hybridization between original and shadow bands. Finally, we assess the distance between the two features to be 2$\Delta_\mathrm{CDW}\sim0.74$\,eV.

\subsubsection{Raman scattering measurements}
\begin{figure}[htb!]
  \centering
  \includegraphics[width=1.0\columnwidth]{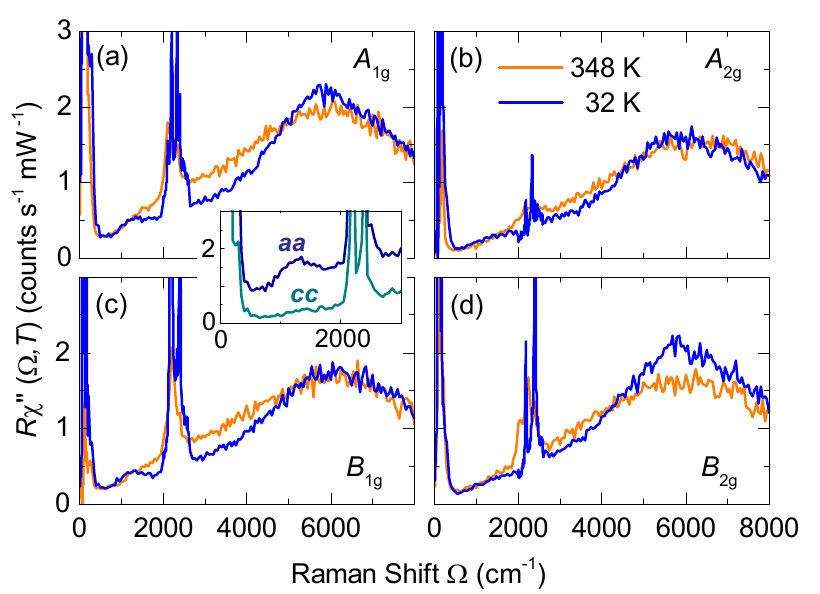}
  \caption{(Color online) Symmetry-resolved Raman spectra of CeTe$_3$ at temperatures as indicated. The excitations in the range 2000 to 2500\,cm$^{-1}$ originate in transitions in the $F_{7/2}$ manifold. They are expected and appear also in $A_{2g}$ symmetry. In all cases they are stronger at low temperature. Gap-like features are predominantly seen in  $A_{1g}$, $B_{1g}$, and $B_{2g}$ symmetries. The inset compares the raw data in $aa$ and $cc$ polarization indicating an $a-c$ anisotropy.
  }
  \label{fig:Raman}
\end{figure}

Polarized Raman spectra of CeTe$_3$ were acquired on a cleaved $(010)$ ($ac$) plane in the temperature range between 348 and 32\,K. Figure~\ref{fig:Raman} shows the symmetry-resolved spectra for 348 and 32\,K. The spectra taken at 348\,K are generally higher by a factor close to 2 and are, therefore, multiplied by 0.55 to make them match those at 32\,K and high energies. Phonons and amplitude modes appear at energies below 400\,cm$^{-1}$.\cite{Lavagnini:2008} In addition, there are also features from crystal-field excitations which appear not only in the Raman active $A_{1g}$, $B_{1g}$ and $B_{2g}$ symmetries but also in $A_{2g}$ symmetry, clearly ruling out a phononic origin. They originate in the energy splitting of the $2F_{7/2}$ state of CeTe$_3$ ($4f$ orbitals) in the spectral region below 400 cm$^{-1}$ and between 2000 and 2500 cm$^{-1}$.

Although $T_{\rm CDW}$ is considerably above room temperature one observes clear changes in the electronic spectra upon cooling. Below 4000-5000\,cm$^{-1}$ the opening of a gap is visible in the $A_{1g}$, $B_{1g}$, and $B_{2g}$ spectra. Around 6000\,cm$^{-1}$ the intensity piles up (Fig.~\ref{fig:Raman}\,(a), (c), (d)). As expected for electronic excitations these temperature-dependent changes are very weak or absent in $A_{2g}$ symmetry (Fig.~\ref{fig:Raman}\,(b)). Therefore we interpret the $A_{1g}$, $B_{1g}$, and $B_{2g}$ spectra in terms of the CDW gap in the electronic excitation spectrum and identify the maximum at $5880\,{\rm cm}^{-1} = 0.73$\,eV, with the maximal gap $2\Delta_\mathrm{CDW}$ as earlier.\cite{Eiter:2013} This energy is in excellent agreement with the gap from ARPES experiments\cite{Brouet:2008} and weak anomalies in the $dI/dV$ curves (see Fig.~\ref{fig_5}) allow us to associate them with the gap. Since the gap depends strongly on momentum\cite{Brouet:2008} no pronounced features can be expected in angle integrated experiments such as STS.

Unexpectedly, we also observe an additional maximum in the $1,200\,{\rm cm}^{-1} \sim 0.15$\,eV range in $A_{1g}$ and $B_{1g}$ which appears only below approximately 200\,K. The appearance in $A_{1g}$ and $B_{1g}$ indicates that the maximum belongs to the lower symmetry of the ordered CDW phase. The comparison of the spectra measured with parallel polarizations along the crystallographic $a$ and $c$ axes (see inset of Fig.~\ref{fig:Raman}) shows that the crystal is mono-domain in the region of the laser spot. The selection rules are exactly those observed before in ErTe$_3$\cite{Eiter:2013}. However, there is no indication of a second CDW in other experiments (see, e.g., Ref. \onlinecite{Brouet:2008}). In addition, there is no direct correspondence to features in the STS results (see asterisks in Fig.~\ref{fig_5}) in that the energy observed by light scattering is approximately a factor of two smaller than expected from tunneling. Speculatively, one may associate the second maximum found in the $aa$-polarized Raman spectra with a hidden transition having a marginally different ordering wave vector and a slightly lower minimum of the free energy, as suggested in Refs.~\onlinecite{Lavagnini:2010,Hu:2009}. However, the selection rules are more indicative of an ordering wave vector orthogonal to the first one (see gapped $cc$ spectrum \textit{versus} peaked $aa$ spectrum in the inset of Fig.~\ref{fig:Raman}) and argue otherwise. Hence, the maximum at $1,200\,{\rm cm}^{-1}$ cannot be explained.


\section{Summary}

Scanning tunneling microscopy and spectroscopy were used for an investigation of the CDW state in CeTe$_3$. The data was acquired at room temperature. Using Raman spectroscopy we confirmed that CeTe$_3$ is in the CDW state at room temperature. The STM topography images clearly show the presence of the CDW modulation. This fact is supported by spectroscopy curves, which indicate that CDW band gap has a value of $\sim 0.37$\,eV and are in good agrement with gap value of $\sim 0.365$\,eV obtained from Raman measurements. The analysis of the Fourier transform images showed two main peaks along the direction of the CDW. One representing the dominant CDW modulation wavevector, located at $|{\bf q}|=10.26\,{\rm nm}^{-1}$, and the other one, $|{\bf c}^\ast-{\bf q}|=4.19\,{\rm nm}^{-1}$, interpreted as a consequence of the wavevector mixing effect. The other peaks, which are not related to the Te sheet atoms and Ce subsurface atoms are related both to the mixing effect and possible structural modulations of the Te sheet atoms. The current images indicate periodic patterns of Te sheet atoms grouped in N and V shapes, which were not previously observed by STM.

\section*{Acknowledgement}
This work was supported by the Serbian Ministry of Education, Science and Technological Development under Projects ON171032, III45018, OI171005. We acknowledge support by the DAAD through the bilateral project between Serbia and Germany (grant numbers 56267076 and 57142964) and the DFG via the Transregional Collaborative Research Center TRR\,80. The collaboration with Stanford University was supported by the Bavarian Californian Technology Center BaCaTeC (grant-no. A5\,[2012-2]). Work in the SIMES at Stanford University and SLAC was supported by the U.S. Department of Energy, Office of Basic Energy Sciences, Division of Materials Sciences and Engineering, under Contract No. DE-AC02-76SF00515.
Part of the work was carried out at the Brookhaven National Laboratory which is operated by the Office of Basic Energy Sciences, U.S. Department of Energy by Brookhaven Science Associates (DE-SC0012704). We are grateful to Nenad Stojilovi\'c for useful discussions.

%

\end{document}